\documentclass[12pt]{iopart}

\expandafter\let\csname equation*\endcsname\relax
\expandafter\let\csname endequation*\endcsname\relax
\usepackage{amsmath}
\usepackage{graphicx}
\usepackage{color}

\newcommand{\p}{\partial}
\newcommand{\f}[2]{\frac{#1}{#2}}

\newcommand{\pd}[2]{\frac{\p #1}{\p #2}}

\newcommand{\parl}{\parallel}
\newcommand{\unit}[1]{\hat{\bm{#1}}}

\makeatletter
\newcommand{\vast}{\bBigg@{4}}
\newcommand{\Vast}{\bBigg@{5}}

\newcommand{\bm}[1]{{\boldsymbol {\mathrm #1}}} % non-italic bold
\newcommand{\bhat}{\unit{b}}

\newcommand{\vth}{v_{\textnormal{th}}}

\newcommand{\vthi}{v_{\textnormal{th},\rmi}}

\newcommand{\vths}{v_{\textnormal{th},s}}
\newcommand{\rhoi}{\rho_{\rmi}}
\newcommand{\tld}{\tilde}
\newcommand{\stella}{\texttt{stella}~}

\newcommand{\beq}{\begin{equation}}
\newcommand{\eeq}{\end{equation}}
\newcommand{\vpa}{u}
\newcommand{\vpas}{u}
\newcommand{\vpe}{v_{\perp}}
\newcommand{\grad}{\nabla}
\newcommand{\mbf}{\mathbf}
\newcommand{\gyroR}[1]{\left<#1\right>_{\mbf{R}}}

\newcommand{\gperp}{\grad_{\perp}}
\newcommand{\phitb}{\varphi}
\newcommand{\vvol}{d^3v}

\newcommand{\defeq}{\doteq}

\newcommand{\bess}{J_0(a_{k,s})}
\newcommand{\bessone}{\frac{J_1(a_{k,s})}{a_{k,s}}}
\newcommand{\gamzero}{\Gamma_0(b_{k,s})}

\newcommand{\gx}[1]{\hat{g}_{#1}}
\newcommand{\phix}[1]{\hat{\varphi}_{#1}}
\newcommand{\gk}[1]{g_{#1\mbf{k}}}
\newcommand{\phik}[1]{\varphi_{#1,\mbf{k}}}
\newcommand{\flxavg}[1]{\left<#1\right>_{\psi}}
\newcommand{\gswap}{g^{\leftrightarrow}_{1s}(k_{\psi},k_{\alpha},\theta,\vpa,\mu,t)}
\newcommand{\phiswap}{\phitb^{\leftrightarrow}_{1}(k_{\psi},k_{\alpha},\theta,t)}
\makeatother
\begin{document}

\title[Intrinsic rotation driven by turbulent acceleration]{Intrinsic rotation driven by turbulent acceleration}

\author{M. Barnes$^{1,2}$ and F. I. Parra$^{1,2}$}

\address{$^1$ Rudolf Peierls Centre for Theoretical Physics, University of Oxford, Clarendon Laboratory, Parks Road, Oxford OX2 3PU, UK}
\address{$^2$ Culham Centre for Fusion Energy, Abingdon OX14 3EA, UK}
\ead{michael.barnes@physics.ox.ac.uk}

\begin{abstract}
Differential rotation is induced in tokamak plasmas when an underlying symmetry of the governing gyrokinetic-Maxwell system of equations is broken.  One such symmetry-breaking mechanism is considered here: the turbulent acceleration of particles along the mean magnetic field.  This effect, often referred to as the `parallel nonlinearity', has been implemented in the $\delta f$ gyrokinetic code \stella and used to study the dependence of turbulent momentum transport on the plasma size and on the strength of the turbulence drive.  For JET-like parameters with a wide range of driving temperature gradients, the momentum transport induced by the inclusion of turbulent acceleration is similar to or smaller than the ratio of the ion Larmor radius to the plasma minor radius.  This low level of momentum transport is explained by demonstrating an additional symmetry that prohibits momentum transport when the turbulence is driven far above marginal stability.
\end{abstract}

\vspace{2pc}
\noindent{\it Keywords}: intrinsic rotation, gyrokinetics, tokamak, turbulence
%\maketitle

\section{Introduction}

Observational evidence obtained from a wide range of tokamaks indicates that axisymmetric plasmas exhibit differential toroidal rotation even in the absence of an externally applied torque (cf.~\cite{riceNF99,riceNF05,bortolonPRL06,scarabosioPPCF06,degrassiePoP07,duvalPPCF07,riceNF07,erikssonPPCF09,incecushmanPRL09,solomonPoP10,parraPRL12}).  This `intrinsic rotation' is determined by momentum redistribution within the plasma, which is typically dominated by turbulent transport.  Understanding turbulent momentum transport is thus critical for predicting intrinsic rotation.

Calculation of the intrinsic turbulent momentum transport in tokamak plasmas is particularly challenging.  This is the result of a symmetry of the gyrokinetic-Maxwell system of equations that statistically prohibits momentum transport to lowest order in the gyrokinetic expansion parameter $\rho_*\defeq \rhoi/a$, with $\rhoi$ the ion Larmor radius and $a$ the plasma minor radius~\cite{peetersPoP05,parraPoP11,sugamaPPCF11,parraPPCF15}.  The symmetry is broken by various physics effects that are formally small in $\rho_*$ and thus neglected in standard $\delta f$ gyrokinetic simulations.  A comprehensive theory including all of these symmetry-breaking mechanisms is given in~\cite{parraNF11,parraPoP12,calvoPPCF12,parraPPCF15,calvoPPCF15}.  There have also been a number of studies dedicated to individual mechanisms, including the effect of diamagnetic flows~\cite{barnesPRL13,leePoP14,leeNF14,leePPCF15,hornsbyNF17}, up-down asymmetry of flux surfaces~\cite{camenenPRL09,camenenPoP09,camenenPRL10,camenenPPCF10,ballPPCF14,ballPPCF16a,ballPPCF16b,ballPPCF16c,ballPPCF17,ballNF17}, slow poloidal variation of fluctuations~\cite{sungPoP13}, and `global' effects~\cite{waltzPoP11,camenenNF11,griersonPRL17,hornsbyNF18}, which include radial profile variation mingled with the other effects mentioned.  Here we consider the effect of turbulent particle acceleration along the mean magnetic field.  The impact of turbulent particle acceleration on intrinsic momentum transport\footnote{The effect of turbulent acceleration on turbulent fluctuations has previously been considered~\cite{kniepCPC04,linJPCS05} and ultimately was shown to be small in $\rho_*$~\cite{candyPoP06b}, as expected from the gyrokinetic orderings introduced in Sec.~\ref{sec:fullsymmetry}} has thus far only been considered for cylindrical magnetic geometry and using a quasilinear analysis~\cite{mcdevittPRL09,mcdevittPoP09}.  Under these conditions finite magnetic shear and/or $E\times B$ shear are necessary to obtain finite momentum transport.  We show that toroidicity provides an additional means for breaking the symmetry of the system, and we supplement our analysis with data from nonlinear gyrokinetic simulations.

%\Pi_{gB}$~\cite{parraPoP11}, with the gyro-Bohm flux $\Pi_{gB} \defeq \rho_*^2 p R$, $p$ the total plasma pressure, and $R$ the plasma major radius. 

With the exception of up-down asymmetry of flux surfaces, all of the symmetry-breaking mechanisms drive momentum transport proportional to $\rho_*$.  In the absence of additional scaling factors to increase the size of the momentum transport, the intrinsic rotation itself is thus a factor of $\rho_*$ smaller than the sonic rotation -- making it dynamically unimportant.  However, as shown in~\cite{parraPPCF15}, the intrinsic momentum transport arising from the various symmetry-breaking mechanisms is theoretically expected to scale with additional factors such as the driving gradients and the ratio of the total to poloidal magnetic field strength, $B/B_p$.  In particular, neoclassical flow effects and finite-orbit-width effects drive turbulent momentum transport of size $\Pi_{\textnormal{int}}/\Pi_{\textnormal{gB}} \sim (k_{\perp}\rho_i)(B/B_p)\rho_*$, with $\Pi_{int}$ the radial component of the toroidal angular momentum flux due to symmetry-breaking, $k_{\perp}$ the characteristic wavenumber of the turbulence in the plane perpendicular to the mean magnetic field, $\Pi_{gB} \defeq \rho_*^2 p R$, $p$ the total plasma pressure, and $R$ the plasma major radius.  The remaining effects -- slow poloidal variation of turbulence, radial profile variation, and turbulent acceleration -- drive turbulent momentum transport of size $\Pi_{\textnormal{int}}/\Pi_{\textnormal{gB}} \sim (k_{\perp}\rho_i)^{-2} \rho_*$.  When turbulent eddies are sufficiently large, i.e., $k_{\perp}\rho_i \sim B_p/B$, all symmetry-breaking mechanisms are the same size.  In principle, this may make it possible to drive intrinsic rotation at levels that, while still sub-sonic, can stabilize MHD modes and potentially suppress turbulence.

In this paper, we use the local, $\delta f$ gyrokinetic code \stella~\cite{barnesJCP18} to simulate electrostatic plasma turbulence, including the effect of turbulent particle acceleration (often referred to as the parallel nonlinearity).  Both $\rho_*$ and the driving temperature gradients are varied in order to determine the scalings of the intrinsic momentum flux and to thus determine the significance of turbulent acceleration in driving intrinsic rotation.  Our results are compared with the theoretical scalings provided in~\cite{parraPPCF15}, and discrepancies are explained via an additional approximate symmetry satisfied by the fluctuations far above marginal stability.

The paper is organised as follows.  In Sec.~\ref{sec:fullsymmetry} we introduce the gyrokinetic-Poisson system of equations and the associated symmetry that prohibits momentum transport.  We then discuss turbulent acceleration and show how it breaks the symmetry of the equations in Sec.~\ref{sec:symbreaking}.  We provide simple scalings for the intrinsic momentum flux due to this symmetry-breaking in Sec.~\ref{sec:scalings} before arguing
for the existence of an additional, approximate symmetry satisfied by the system in Sec.~\ref{sec:reducedsymmetry}.  Numerical results are presented in Sec.~\ref{sec:results}, and a summary with discussion of implications is given in Sec.~\ref{sec:conclusions}.

\section{Symmetry of the gyrokinetic-Poisson system}\label{sec:fullsymmetry}

Low-frequency fluctuations in tokamak plasmas are described by the gyrokinetic-Maxwell system of equations~\cite{cattoPP78,friemanPoF82,brizardRMP07,parraPPCF08,parraPPCF11,abelRPP13}.  They are obtained by averaging over particle gyration about the mean magnetic field, with the assumption that fluctuations evolve on a much longer time scale than the gyration period.  If one further assumes a space-time scale separation between the fluctuations and the mean plasma profiles, then one obtains the local, $\delta f$ gyrokinetic model.  Explicitly, we restrict our attention to electrostatic fluctuations and impose the ordering
\beq
\frac{\delta f_s}{f_s} \sim \frac{\omega}{\Omega_s} \sim \frac{\rho_s}{L} \sim  \frac{k_{\parallel}}{k_{\perp}} \sim k_{\parallel}\rho_s \sim \frac{e\phix{}}{T_s}\sim  \epsilon \ll 1,
\label{eqn:GKordering}
\eeq
where $\epsilon$ is the fundamental gyrokinetic expansion parameter, $f_s=F_s+\delta f_s$ is the particle distribution function for species $s$, $F_s$ and $\delta f_s$ are its mean and fluctuating components, $\phix{}$ is the electrostatic potential, $\omega$ is a characteristic fluctuation frequency, $\Omega_s=Z_s eB/m_sc$ is the Larmor frequency, $Z_s$ is particle charge number, $m_s$ is particle mass, $c$ is the speed of light, $e$ is the proton charge, $B$ is the magnetic field strength, $\rho_s=\vths/\Omega_s$ is the thermal Larmor radius, $\vths=\sqrt{2T_s/m_s}$, $T_s$ is temperature, $L$ is a characteristic length associated with mean plasma profiles, and $k_{\parallel}$ and $k_{\perp}$ are characteristic fluctuation wavenumbers along and across the mean magnetic field.

Gyro-averaging the Fokker-Planck equation, applying the gyrokinetic ordering~(\ref{eqn:GKordering}), and expanding $f=f_0 + f_1+f_2+...$, with $f_\alpha =\Or(\epsilon^\alpha) f$, yields a gyrokinetic equation describing the evolution of $\gx{}$, the distribution of particle guiding centres.  We choose to work in $(\mbf{R},\vpa,\mu,\vartheta)$ coordinates, with $\mbf{R}$ the particle guiding centre position, $\mu=mv_{\perp}^2/2B$ the lowest order particle magnetic moment, $\vpa$ the particle velocity along the magnetic field, $\vpe$ the particle speed across the magnetic field, respectively, and $\vartheta$ the particle gyrophase.  In these coordinates the gyrokinetic equation valid to lowest order in $\epsilon$ is
\beq
\begin{split}
\pd{\gx{1s}}{t}&+\vpas\bhat\cdot\left(\grad\gx{1s}+\f{Z_se}{T_s}F_{0s}\grad\gyroR{\phix{1}}\right) 
+\mbf{v}_{Ms}\cdot\left(\gperp \gx{1s} + \f{Z_s e}{T_s}F_{0s} \gperp \gyroR{\phix{1}}\right) \\
&+\dot{u}_{0s}\pd{\gx{1s}}{\vpas}
+ \f{c}{B}\{\gyroR{\phix{1}},\gx{1s}\}
+\gyroR{\mbf{v}_{E1}}\cdot\grad\big|_E F_{0s}= \hat{C}[\gx{1s}],
\end{split}
\label{eqn:gkeLOx}
\eeq
where $\gyroR{.}$ denotes a gyro-average at fixed guiding centre position $\mbf{R}$, $\phix{1}$ is the electrostatic potential generated by $\gx{1}$, $t$ is time, $\bhat$ is the unit vector along the mean magnetic field, $F_{0s}$ is taken to be a Maxwellian distribution in particle velocity, $\dot{u}_{0s}=-(\mu/m_s)\unit{b}\cdot\grad B$ is the lowest order contribution to the parallel acceleration, $\grad|_E$ is a gradient taken at fixed particle kinetic energy $E=m\vpa^2/2+\mu B$, $\mbf{v}_{E1}=(c/B)\bhat\times\nabla_{\perp}\phix{1}$ is the $E\times B$ drift velocity, $\{.,.\}$ is a Poisson bracket, $\mbf{v}_{Ms}=(\bhat/\Omega_s)\times\left(\mu\grad B+\vpa^2\bm{\kappa}\right)$, $\bm{\kappa}=\bhat\cdot\grad\bhat$, and the operator $\hat{C}$ accounts for the effect of collisions on $\gx{1}$.  The system is closed by coupling to Poisson's equation, which reduces to quasineutrality when the Debye length is much smaller than the electron Larmor radius:
\beq
\sum_s Z_s e \int \vvol \left(\gx{1s} + \frac{Z_s e}{T_s}\left(\gyroR{\phix{1}}-\phix{1}\right)F_{0s}\right) = 0.
\label{eqn:QNLOx}
\eeq

It will be convenient for much of the paper to work in Fourier space, so we define the Fourier components of $\gx{}$ via $\gk{}\defeq\mathcal{F}_{\mbf{k}}[\gx{}]$, with $\mathcal{F}_{\mbf{k}}$ denoting the two-dimensional, discrete Fourier transform in the plane perpendicular to $\bhat$ and $\mbf{k}$ denoting the wave vector in this plane.  We use the coordinate system $(\alpha,\psi,\theta)$ to represent physical space, with $\psi$ a flux surface label, $\alpha$ a field line label, and $\theta$ a poloidal angle measuring distance along a given magnetic field line.
Applying $\mathcal{F}_{\mbf{k}}$ to~(\ref{eqn:gkeLOx}) and~(\ref{eqn:QNLOx}) gives
\beq
\begin{split}
\pd{\gk{1s,}}{t}&+\vpas\bhat\cdot\grad \theta\left(\pd{\gk{1s,}}{\theta}+\f{Z_se}{T_s}\pd{\bess\phik{1}}{\theta}F_{0s}\right) 
+i\mbf{v}_{Ms}\cdot\mbf{k}\left(\gk{1s,} + \f{Z_s e}{T_s}F_{0s} \bess\phik{1}\right) \\
&+\dot{u}_{0s}\pd{\gk{1s,}}{\vpas}
+ \f{c}{B}\mathcal{F}_{\mbf{k}}\left[\{\gyroR{\phix{1}},\gx{1s}\}\right]
+ik_{\alpha}c\bess\phik{1}\pd{F_{0s}}{\psi}\bigg|_{E}= C_{\mbf{k}}[\gk{1s,}],
\end{split}
\label{eqn:gkeLOk}
\eeq
and
\beq
\sum_s Z_s e \left(\int \vvol \bess\gk{1s} + \frac{Z_s e n_s}{T_s}\left(\gamzero-1\right)\phik{1}\right) = 0,
\label{eqn:QNLOk}
\eeq
where $n_s$ is the plasma density, $J_0$ is a Bessel function of the first kind, $a_{k,s}=k\vpe/\Omega_s$, $\Gamma_0(b)=\exp(-b)I_0(b)$, $I_0$ is a modified Bessel function of the first kind, $b_{k,s}=k^2\rho_s^2/2$, and $C_{\mbf{k}}[\gk{1s,}]\defeq \mathcal{F}_{\mbf{k}}[\hat{C}[\gx{1s}]]$.

If the confining magnetic geometry is up-down symmetric, the gyrokinetic-Poisson system~(\ref{eqn:gkeLOk}) and~(\ref{eqn:QNLOk}) possesses a symmetry that inhibits momentum transport: If $g_{1s}(k_{\psi},k_{\alpha},\theta,\vpa,\mu,t)$ is a solution with associated potential $\varphi_1(k_{\psi},k_{\alpha},\theta,t)$, then $\gswap = -g_{1s}(-k_{\psi},k_{\alpha},-\theta,-\vpa,\mu,t)$ is also a solution with associated potential $\phiswap=-\varphi_1(-k_{\psi},k_{\alpha},-\theta,t)$~\cite{peetersPoP05,parraPoP11,sugamaPPCF11,parraPPCF15}.  For turbulence in a statistical steady state that is independent of initial conditions, $g_{1s}$ and $g_{1s}^{\leftrightarrow}$ occur with equal frequency.  Upon statistical average, this leads to a vanishing lowest-order, radial transport of toroidal angular momentum $\Pi_1=\flxavg{|\grad\psi|}^{-1}\int d^3v \flxavg{\left(m R^2\delta f\mbf{v} \cdot\nabla\zeta \right)\left(\mbf{v}_E\cdot \grad\psi\right)}$, where $\zeta$ is toroidal angle, $\flxavg{A}\defeq (\int d\zeta d\theta \mathcal{J})^{-1}\int d\zeta d\theta \mathcal{J} A$ denotes an average over the flux surface, and $\mathcal{J}=\mathbf{B}\cdot\nabla\theta$ is the Jacobian of the transform to $(\zeta,\psi,\theta)$ coordinates.  The statistical average could be a time average over many nonlinear decorrelation times in a statistical steady state or an ensemble average over many turbulence realisations.  We use the former definition in the simulation results that follow.  The fact that $\Pi_1$ vanishes can be deduced by examining the contribution to $\Pi_1$ from wavevector $\mbf{k}$, given by
\beq
\begin{split}
\Pi_{1,\mbf{k}} &= -\frac{1}{\flxavg{|\grad\psi|}} \sum_s \flxavg{\frac{m_sc}{B} k_{\alpha} \phik{1}^* 
\int\vvol \gk{1s,}\left(i\vpas I(\psi) \bess + \mbf{k}\cdot\grad\psi \frac{\vpe^2}{\Omega_s} \bessone\right)} \\
& -\frac{1}{\flxavg{|\grad\psi|}} \sum_s \flxavg{\f{m_s n_s c^2}{B^2}k_{\alpha} \mbf{k}\cdot\grad\psi \left|\phik{1}\right|^2 
\left(\Gamma_0(b_{\mbf{k},s})-\Gamma_1(b_{\mbf{k},s})\right)},
\end{split}
\eeq
with $I(\psi)=RB_{\zeta}$, $R$ the plasma major radius, $B_{\zeta}$ the toroidal component of the magnetic field, $\Gamma_1(b)=\exp(-b)I_1(b)$, and $*$ denoting complex conjugation.  Applying the symmetry discussed above, we see that the lowest order contribution to the radial flux of toroidal angular momentum, $\Pi_1=\sum_{\mbf{k}}\overline{\Pi_{1,\mbf{k}}}$, is zero, with the overline denoting a statistical average.

\section{Symmetry-breaking induced by turbulent acceleration}\label{sec:symbreaking}

The symmetry of the lowest order gyrokinetic equation~(\ref{eqn:gkeLOk}) is broken when one takes into account various physics effects that are formally small in the gyrokinetic expansion parameter $\epsilon$~\cite{parraNF11,parraPPCF15}.  Here we focus on one such symmetry-breaking mechanism, the turbulent parallel acceleration of particles.  Retaining higher order terms, the force parallel to the mean magnetic field is given by
\beq
m_s\dot{u}_s = -\left(\unit{b}+\frac{\vpas}{\Omega_s}\unit{b}\times\bm{\kappa}\right)\cdot\left(\mu\grad B + Z_s e\grad \phix{}\right) + \mathcal{O}\left(\rho_{*s}^2\f{T_s}{a}\right),
\eeq
with $a$ the minor radius of the plasma volume and $\rho_{*s}=\rho_s/a$.
Defining $\dot{u}_{1s}=\dot{u}_s-\dot{u}_{0s}+\mathcal{O}(\rho_{*s}^2\vths^2/a)$, we have
\beq
m_s\dot{u}_{1s} = -Z_s e \unit{b}\cdot\grad\phix{1} - \frac{\vpas}{\Omega_s}\unit{b}\times\bm{\kappa}\cdot\left(\mu\grad B + Z_s e\grad \phix{1}\right).
\label{eqn:u1dot}
\eeq
We see that, in contrast to the lowest order parallel acceleration $\dot{u}_{0}$, the acceleration $\dot{u}_1$ is turbulent in nature; i.e., it depends on the fluctuating electrostatic potential $\phix{1}$.
The second term in~(\ref{eqn:u1dot}) is the only one independent of turbulence amplitude, and it can be manipulated into the form $(\mu\vpas/\Omega_s)\unit{b}\times\bm{\kappa}\cdot\grad B = \beta' (\mu\vpas/\Omega_s)I\unit{b}\cdot\grad B$, with $\beta'\defeq (4\pi/B^2)\partial p_{\textnormal{tot}}/\partial \psi$ and $p_{\textnormal{tot}}$ the total plasma pressure.  As the plasma pressure in tokamaks is small compared to the magnetic pressure, $|\beta'|$ is typically small.  The parallel acceleration given by~(\ref{eqn:u1dot}) is then dominated by the turbulent contributions.

The breaking of symmetry induced by inclusion of $\dot{u}_1$ can be seen by comparing how $\dot{u}_0$ and $\dot{u}_1$ behave under the transformation ($\theta\rightarrow-\theta$, $\vpa\rightarrow-\vpa$, $k_{\psi}\rightarrow-k_{\psi}$).  We see that $\dot{u}_0(\theta,\vpa,k_{\psi})=-\dot{u}_0(-\theta,-\vpa,-k_{\psi})$, while $\dot{u}_1(\theta,\vpa,k_{\psi})=\dot{u}_1(-\theta,-\vpa,-k_{\psi})$.  This difference in parity mars the symmetry described in Sec.~\ref{sec:fullsymmetry} and leads to finite steady-state momentum transport.  Replacing $\gk{1s,}$ with $\gk{s,}$ and $\dot{u}_{0s}$ with $\dot{u}_s$ in~(\ref{eqn:gkeLOk}) and~(\ref{eqn:QNLOk}), and defining $\gk{2s,}\defeq\gk{s,}-\gk{1s,}$, the gyrokinetic-Poisson system becomes
\beq
\begin{split}
\pd{\gk{2s,}}{t}&+\vpas\bhat\cdot\grad \theta\left(\pd{\gk{2s,}}{\theta}+\f{Z_se}{T_s}\pd{\bess\phik{2}}{\theta}F_{0s}\right) 
+i\mbf{v}_{Ms}\cdot\mbf{k}\left(\gk{2s,} + \f{Z_s e}{T_s}F_{0s} \bess\phik{2}\right) \\
&+\dot{u}_{0s}\pd{\gk{2s,}}{\vpas} + \mathcal{F}_{\mbf{k}}\left[\dot{u}_{1s}\pd{\gx{s}}{\vpas}
+ \f{c}{B}\left(\{\gyroR{\phix{}},\gx{s}\}-\{\gyroR{\phix{1}},\gx{1s}\}\right)\right]\\
&+ik_{\alpha}c\bess\phik{2}\pd{F_{0s}}{\psi}\bigg|_{E}= C_{\mbf{k}}[\gk{2s,}]
\end{split}
\label{eqn:gkeHOk}
\eeq
and
\beq
\sum_s Z_s e \left(\int \vvol \bess\gk{2s} + \frac{Z_s e n_s}{T_s}\left(\gamzero-1\right)\phik{2}\right) = 0.
\label{eqn:QNHOk}
\eeq
%\beq
%\begin{split}
%\pd{\gk{2s,}}{t}&+\vpas\bhat\cdot\grad \theta\left(\pd{\gk{2s,}}{\theta}+\f{Z_se}{T_s}\pd{\bess\phik{2}}{\theta}F_{0s}\right) 
%+i\mbf{v}_{Ms}\cdot\mbf{k}\left(\gk{2s,} + \f{Z_s e}{T_s}F_{0s} \bess\phik{2}\right) \\
%&+\dot{u}_{0s}\pd{\gk{2s,}}{\vpas} + \mathcal{F}_{\mbf{k}}\left[\dot{u}_{1s}\pd{\gx{1}}{\vpas}
%+ \f{c}{B}\left(\{\gyroR{\phix{2}},\gx{1s}\}+\{\gyroR{\phix{1}},\gx{2s}\}\right)\right]\\
%&+ik_{\alpha}c\bess\phik{2}\pd{F_{0s}}{\psi}\bigg|_{E}= C_{\mbf{k}}[\gk{2s,}]
%\end{split}
%\label{eqn:gkeHOk}
%\eeq
%and
%\beq
%\sum_s Z_s e \left(\int \vvol \bess\gk{2s} + \frac{Z_s e n_s}{T_s}\left(\gamzero-1\right)\phik{2}\right) = 0,
%\label{eqn:QNHOk}
%\eeq

%  As a result~(\ref{eqn:gkeHOk}) and~(\ref{eqn:QNHOk}) possess a symmetry opposite to that of~(\ref{eqn:gkeLOk}) and~(\ref{eqn:QNLOk}): If $g_2(k_{\psi},k_{\alpha},\theta,\vpa,\mu,t)$ is a solution with associated potential $\varphi_2(k_{\psi},k_{\alpha},\theta,t)$, then there is another solution $g_2^{\leftrightarrow}(k_{\psi},k_{\alpha},\theta,\vpa,\mu,t)=g_2(-k_{\psi},k_{\alpha},-\theta,-\vpa,\mu,t)$ with associated potential $\varphi_2^{\leftrightarrow}(k_{\psi},k_{\alpha},\theta,t)=\varphi_2(-k_{\psi},k_{\alpha},-\theta,t)$.  This makes the next order corrections to the particle and energy fluxes vanish, while the next order correction to the momentum flux, $\sum_{\mbf{k}}\Pi_{2,\mbf{k}}$, is non-zero.  The expression for $\Pi_{2,\mbf{k}}$ is 

For $\gk{2s,} \ll \gk{1s,}$, the product of $\gk{2s,}$ and $\phik{2}$ can be neglected when calculating the radial flux of toroidal angular momentum.  The resulting expression for the lowest order (non-vanishing) momentum flux is $\Pi_2 = \sum_{\mbf{k}} \Pi_{2,\mbf{k}}$, with
\beq
\begin{split}
\Pi_{2,\mbf{k}} &= -\frac{1}{\flxavg{|\grad\psi|}} \sum_s \flxavg{\frac{m_sc}{B} k_{\alpha} \phik{2}^* 
\int\vvol \gk{1s,}\left(\rmi\vpas I(\psi) \bess + \mbf{k}\cdot\grad\psi \frac{\vpe^2}{\Omega_s} \bessone\right)} \\
&-\frac{1}{\flxavg{|\grad\psi|}} \sum_s \flxavg{\frac{m_sc}{B} k_{\alpha} \phik{1}^* 
\int\vvol \gk{2s,}\left(\rmi\vpas I(\psi) \bess + \mbf{k}\cdot\grad\psi \frac{\vpe^2}{\Omega_s} \bessone\right)} \\
& -\frac{2}{\flxavg{|\grad\psi|}} \sum_s \flxavg{\f{m_s n_s c^2}{B^2}k_{\alpha} \mbf{k}\cdot\grad\psi \textnormal{Re}[\phik{1}^*\phik{2}]
\left(\Gamma_0(b_{\mbf{k},s})-\Gamma_1(b_{\mbf{k},s})\right)},
\label{eqn:PiHO}
\end{split}
\eeq
where $\textnormal{Re}[.]$ denotes the real part.

\section{Momentum flux scalings}\label{sec:scalings}

We are interested in determining how the amplitude of the momentum flux scales with quantities such as device size, eddy size, and driving gradients. The expected amplitude of the momentum flux given by~(\ref{eqn:PiHO}) depends on the fluctuation
amplitudes and wavenumbers, as well as the phases between different fluctuations.  To obtain the aforementioned scalings for the momentum flux, we must thus first deduce the scalings for the fluctuations.  To do this we make a number of assumptions along the lines of Refs.~\cite{barnesPRL11b} and~\cite{parraPPCF15}, where similar scalings for turbulent heat and momentum fluxes are obtained.  In particular, we assume: that phase differences between $\gk{}$ and $\phitb_{\mbf{k}}$ lead to no more than order unity variations in the flux; that the fluctuations are isotropic in the plane perpendicular to the mean magnetic field so that $k_{\alpha}\alpha \sim k_{\psi}\psi \sim k_{\perp}\rhoi$; that at the outer scale the nonlinear transfer rate $\tau_{\mbf{k}}^{-1}$ is comparable to the energy injection rate, which we estimate to be of order $k_{\perp}\rhoi\vth/L_T$, with $L_T$ the ion temperature gradient scale length; and that the plasma is in a state of critical balance~\cite{goldreichApJ95} so that the time scale associated with parallel propagation $(k_{\parl}\vth)^{-1}$ is comparable to the nonlinear turnover time $\tau_{\mbf{k}}$ at all spatial scales.

Assuming $e\phik{1}/T \sim \gk{1s,}/F_{0s}$ and $\tau_{\mbf{k}}^{-1} \sim (\mbf{v}_E\cdot \grad)_{\mbf{k}}\sim(k_{\perp}\rhoi)^2 (\vth/\rhoi)(e\phik{1}/T)$, we obtain
\beq
k_{\parl}\vth \sim \left(k_y\rhoi\right)^2\frac{\vth}{\rhoi}\frac{e\phik{1}}{T}\sim k_{\perp}\rhoi\frac{\vth}{L_T},
\eeq
where we have taken $a\sim L_n \sim L_T$, with $L_n$ the density gradient scale length.  Balancing the first and last terms gives $k_{\parl}L_T \sim k_{\perp}\rhoi$, and balancing the last two terms gives $e\phik{1}/T\sim (k_{\perp}L_T)^{-1}$.  If $k_{\parl}$ is set by the system size, then these scalings predict that the characteristic $k_{\perp}$ of the turbulence decreases and that the fluctuation amplitudes rapidly increase with increasing temperature gradient.  The same trends are obtained if instead the minimum $k_{\perp}$ is set by linear stability thresholds, which would make the minimum $k_{\perp}$ decrease with increasing temperature gradient.  Gyrokinetic simulations of plasma turbulence far from marginal stability have found results consistent with these predictions~\cite{barnesPRL11b}.

Now that we have a predicted scaling for $\phik{1}$ -- and thus $\gk{1s,}$ -- we proceed to obtain the scaling for $\gk{2s,}$.  We argued above that the time scale associated with the fluctuations is $k_{\perp}\rhoi (\vth/L_T)$.  Using this time scale and balancing $\partial \gk{2s,}/\partial t$ with the source terms containing $\gk{1s,}$ and $\phik{1}$ in~(\ref{eqn:gkeHOk}), we have
\beq
k_{\perp}\rhoi \frac{\vth}{L_T}\gk{2s,} \sim \mathcal{F}_{\mbf{k}}
\left[\dot{u}_{1s}\pd{\gx{1}}{\vpas}\right]
\sim \frac{\vth}{L_T}\frac{\rhoi}{L_T}\gk{1s,} \sim \frac{\vth}{L_T}\left(\frac{\rhoi}{L_T}\right)^2 \frac{1}{k_{\perp}\rhoi}F_{0s},
\eeq
from which we find $\gk{2s,} \sim (k_{\perp}L_T)^{-2}F_{0s}$.  Substituting the
scalings for $\gk{1s,}$, $\gk{2s,}$, $\phik{1}$, and $\phik{2}$ into~(\ref{eqn:PiHO}) gives
\beq
\frac{\Pi_{2,\mbf{k}}}{Q_{1\textnormal{i},\mbf{k}}} \frac{\vth}{R} \sim \frac{\rhoi}{L_T}\frac{1}{k_{\perp}\rhoi},
\label{eqn:pi2scaling}
\eeq
where the lowest-order contribution to the ion radial energy transport is $Q_{1\textnormal{i}} = \sum_{\mbf{k}}Q_{1\textnormal{i},\mbf{k}}$, with
\beq
Q_{1s,\mbf{k}} =  -\frac{\textnormal{i}ck_{\alpha}}{\flxavg{|\grad\psi|}} \flxavg{\phik{1}^* 
\int\vvol \gk{1s,}\bess\left(\frac{m_sv^2}{2} \right)}.
\eeq
Our use of the ion energy flux to normalize $\Pi_{2,\mbf{k}}$ in~(\ref{eqn:pi2scaling}) is motivated by the fact that $\Pi_{1,\mbf{k}}=0$.

The scaling relation~(\ref{eqn:pi2scaling}) implies that the intrinsic momentum flux arising from the turbulent parallel acceleration is always small in the gyrokinetic expansion parameter $\epsilon\sim\rhoi/L$ and is minimum near marginal stability where both $k_{\perp}$ and $L_T$ are relatively large.  However, 
as we discuss in Sec.~\ref{sec:reducedsymmetry}, an additional symmetry of the gyrokinetic-Poisson system may be approximately satisfied when both $k_{\perp}$ and $L_T$ become sufficiently small.  If so, the momentum transport induced by turbulent acceleration could be much smaller than the estimate given by~(\ref{eqn:pi2scaling}).

%as discussed in Sec.~\ref{sec:reducedsymmetry}, turbulent acceleration is likely to induce little to no 
%momentum transport far from marginal stability where an additional symmetry of the gyrokinetic-Poisson system is approximately satisfied.  Consequently, it is anticipated that the momentum flux resulting from parallel acceleration will typically be small, regardless of the strength of the turbulence drive.

\section{Additional symmetry for reduced system}\label{sec:reducedsymmetry}

In a system with no magnetic shear and no magnetic drift in the radial direction, an additional symmetry of the gyrokinetic-Poisson system of equations exists. Namely, if $g_{1s}(k_{\psi},k_{\alpha},\theta,\vpa,\mu,t)$ is a solution with associated potential $\varphi_1(k_{\psi},k_{\alpha},\theta,t)$, then $\gswap=g_{1s}(k_{\psi},k_{\alpha},-\theta,-\vpa,\mu,t)$ is also a solution with associated potential $\phiswap=\varphi_1(k_{\psi},k_{\alpha},-\theta,t)$~\cite{parraPPCF15}.  This differs from the symmetry of the full gyrokinetic-Poisson system in that there is no need to change the sign of the radial wavenumber $k_{\psi}$ and of $\gx{1s}$ and $\phix{1}$.  

While the parallel acceleration $\dot{u}_1$ breaks the full symmetry discussed in Sec.~\ref{sec:fullsymmetry}, it does not break the symmetry of a system with neither magnetic shear nor a radial magnetic drift -- as long as the second term in~(\ref{eqn:u1dot}) can be neglected.  As discussed in Sec.~\ref{sec:symbreaking}, this is a good approximation when $\beta'$ is small or the turbulence amplitude is large.  Because of the additional symmetry of the reduced system, $\varphi_1$ does not change sign when $\theta\rightarrow-\theta$, and so $\dot{u}_1^{\leftrightarrow}(\theta,\vpa)=-\dot{u}_1(-\theta,-\vpa)$.  This sign reversal under the transformation $(\vpa,\theta)\rightarrow(-\vpa,-\theta)$ is identical to the behavior of the lowest order acceleration $\dot{u}_0$ and thus does not break the symmetry of the reduced gyrokinetic-Poisson system.  Consequently, the turbulent acceleration does not contribute to momentum transport.

Although the systems in which we are interested in general have both magnetic shear and a radial magnetic drift, it is still possible for this additional symmetry to be approximately satisfied.  For systems far from marginal stability with $R/L_T \gg 1$, the radial magnetic drift often has only a small effect on linear growth rates and nonlinear physics; an illustrative example is provided in Sec.~\ref{sec:results} (see Fig.~\ref{fig:omega_vs_ky}).  A possible reason for this is the fact that the time scale associated with the radial magnetic drift is small compared to that of the background gradient drive (and thus the streaming and nonlinear turnover times via the critical balance argument of Sec.~\ref{sec:scalings}) by a factor of $R/L_T$.  When the radial magnetic drift is unimportant, the magnetic shear appears in the gyrokinetic-Poisson system only through the perpendicular wavenumber as an argument to the Bessel function.  For turbulence peaked at long wavelengths -- as we argue in Sec.~\ref{sec:scalings} is the case far from marginal stability -- the Bessel function is approximately independent of $k_{\perp}$.  In this limit the magnetic shear  plays little role as well.  It is thus possible that the additional symmetry described here is approximately satisfied as turbulence is driven beyond marginal stability.  Consequently, it is expected that the momentum transport driven by parallel acceleration will be small for turbulence far from marginal stability.

\section{Simulation equations and results}\label{sec:results}

To test the predictions for the size of the momentum flux arising from the inclusion of turbulent acceleration, we have implemented the $\dot{u}_{1s}$ terms given by~(\ref{eqn:u1dot}) in the local, $\delta f$ gyrokinetic code \stella~\cite{barnesJCP18}.  For the sake of simulation efficiency, we do not separately evolve $\gk{1s,}$ and $\gk{2s,};$ instead, we simulate a single equation for $\gk{s,}=\gk{1s,}+\gk{2s,}$, obtained by summing the two lowest order equations~(\ref{eqn:gkeLOk}) and~(\ref{eqn:gkeHOk}):
\beq
\begin{split}
\pd{\gk{s,}}{t}&+\vpas\bhat\cdot\grad \theta\left(\pd{\gk{s,}}{\theta}+\f{Z_se}{T_s}\pd{\bess\varphi_{\mbf{k}}}{\theta}F_{0s}\right) 
+\mathcal{F}_{\mbf{k}}\left[\dot{u}_s\pd{\gx{s}}{\vpas} + \frac{c}{B}\{\gyroR{\phix{}}, \gx{s}\} \right]\\
&+i\mbf{v}_{Ms}\cdot\mbf{k}\left(\gk{s,} + \f{Z_s e}{T_s}F_{0s} \bess\varphi_{\mbf{k}}\right)
+ick_{\alpha}\bess\varphi_{\mbf{k}}\pd{F_{0s}}{\psi}\bigg|_E = C_k[\gk{s,}],
\end{split}
\label{eqn:gke}
\eeq
where the collision operator $C_k$ used in \stella is a gyrokinetic form~\cite{mandellJPP18} of the Dougherty collision operator~\cite{doughertyPoF64}, a Fokker-Planck operator that satisfies Boltzmann's H-Theorem and conserves particle number, momentum and energy.  The associated quasineutrality constraint is identical to~(\ref{eqn:QNLOk}) with the substitution $\gk{1s,}\rightarrow \gk{s,}$.  Note that we have implicitly included a number of terms at even higher order in~(\ref{eqn:gke}) by including products of $\phik{2}$ and $\gk{2s,}$ in the nonlinearities.  These should not affect our results, provided $\epsilon$ is sufficiently small.

\begin{table}[t]
\caption{\label{t:params}Equilibrium plasma parameters for \stella simulations}
\begin{indented}
\item[]\begin{tabular}{@{}lll}
\br
Parameter&Description&Value\\
\mr
$\tld{r}=r/a$& normalized minor radius &0.5 \\
$\tld{R}=R_0/a$& normalized major radius & 3.2 \\
$\rmd R_0/\rmd r$& local Shafranov shift & -0.2 \\
$q$ & safety factor & 1.7 \\
$\hat{s}=\rmd\ln q/\rmd \ln r$& magnetic shear & 0.7\\
$\kappa$& elongation & 1.35 \\
$\rmd\kappa/\rmd \tld{r}$ & elongation derivative & 0.1 \\
$\delta$ & triangularity & 0.1 \\
$\rmd\delta/\rmd \tld{r}$ & triangularity derivative & 0.2 \\
$ (4\pi p_{\textnormal{tot}} / B_r^2) (\rmd \ln p_{\textnormal{tot}}/\rmd\tld{r})$ & normalized $\beta'$ & -0.035 \\
$\nu_{ii} (a/\vthi)$& ion-ion collision frequency & 0.005 \\
$a/L_n$ & inverse density gradient scale length & 0.7 \\
$B_r$ & reference magnetic field strength & $RB_{\zeta}/R_0$\\
\br
\end{tabular}
\end{indented}
\end{table}

For our simulations we use a Miller local specification of the magnetic geometry~\cite{millerPoP98}, in which the cylindrical coordinates $R$ and $Z$ are expressed as $R(r,\theta) = R_0(r) + r \cos(\theta + \sin\theta \arcsin \delta(r))$ and $Z(r,\theta)= \kappa(r) r\sin(\theta)$.  Here $\kappa$ and $\delta$ measure elongation and triangularity of the target flux surface, and $r$ and $R_0$ are averages of the minimum and maximum values of the minor and major radii of the target flux surface at the height of the magnetic axis.  The fixed parameter values used in our \stella simulations, chosen to be similar to those of a typical JET shot at mid-radius, are given in Table~\ref{t:params}.  In order to test our scaling predictions for the intrinsic momentum flux~(\ref{eqn:pi2scaling}), we conducted scans in both $\rho_*$ and $a/L_T$.  These scans are intended to determine the intrinsic momentum flux as a function of plasma volume and distance from marginal stability, respectively.

\begin{figure}[htpb]
  \begin{center}
    \includegraphics*[width=0.49\textwidth]{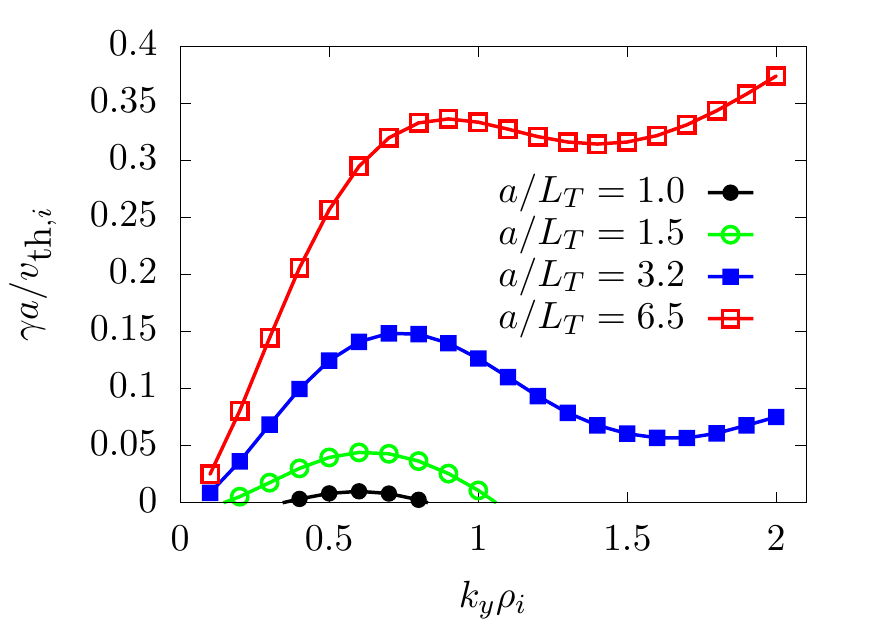}
     \includegraphics*[width=0.49\textwidth]{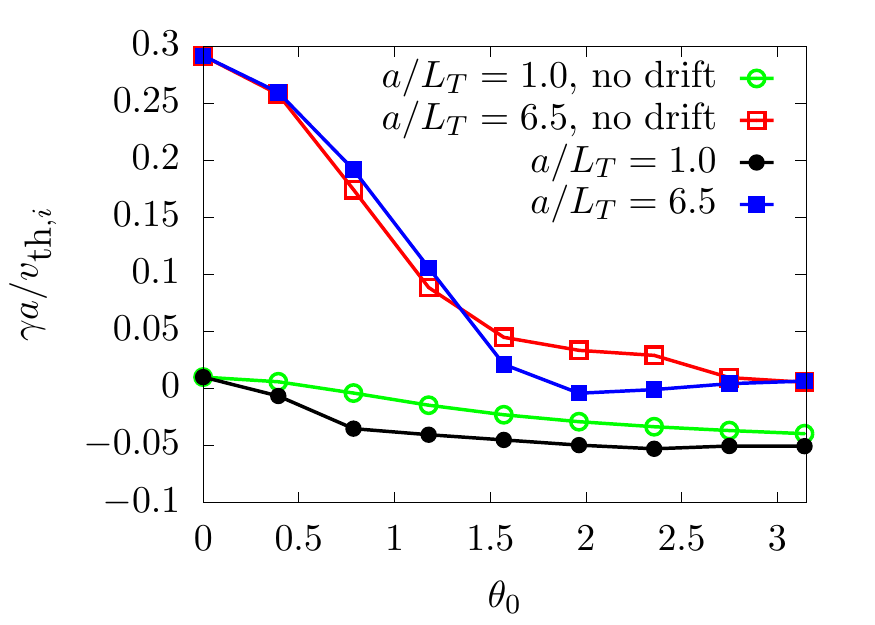}
  \end{center}
  \caption{(Left): Normalized linear growth rate $\gamma$ vs normalized bi-normal wavenumber $k_y\rhoi$ for $k_x=0$ and different values of the equilibrium temperature gradient scale length $L_T$.  (Right): Normalized linear growth rate $\gamma$ vs ballooning angle $\theta_0=k_x/k_y\hat{s}$ for $k_y\rho_i=0.6$, with and without the radial component of the magnetic drift artificially set to zero.}
  \label{fig:omega_vs_ky}
\end{figure}

All simulations discussed here treated electrons and a single deuterium ion species kinetically and used 48 grid points in $\vpa$, 12 grid points in $\mu$, and 32 grid points per $2\pi$ segment in $\theta$.  The results of linear simulations with $\rho_*=0$ and $a/L_T$ varying from 1 to 6.5 are given in Fig.~\ref{fig:omega_vs_ky}.  These simulations used an extended ballooning domain spanning $[-3\pi,3\pi]$ in $\theta$.  We see from the growth rate spectrum that $a/L_T=1$ is very near the linear critical gradient, with only a narrow range of weakly-unstable bi-normal mode numbers ($k_y$).  In contrast, $a/L_T=6.5$ is far above marginal stability, with relatively large growth rates across the entire spectrum and no finite cutoff at long wavelengths.  In the former case, one anticipates that the largest turbulent eddies are determined by the minimum $k_{\perp}$ for which there is a non-zero growth rate; in the latter case, the largest turbulent eddies are constrained by the connection length along the magnetic field via the critical balance argument summarized in Sec.~\ref{sec:scalings}.  This range of $a/L_T$ should thus give a good indication of how the intrinsic momentum flux varies with distance from marginality.  The righthand plot in Fig.~\ref{fig:omega_vs_ky}, which shows the variation in growth rate as a function of the ballooning angle $\theta_0\defeq k_x/(\hat{s}k_y)$, demonstrates the relative unimportance of the radial component of the magnetic drift for calculating the linear growth rate when the system has large $R/L_T$ and is far from marginal stability.

\begin{figure}[htpb]
  \begin{center}
    \includegraphics*[width=0.49\textwidth]{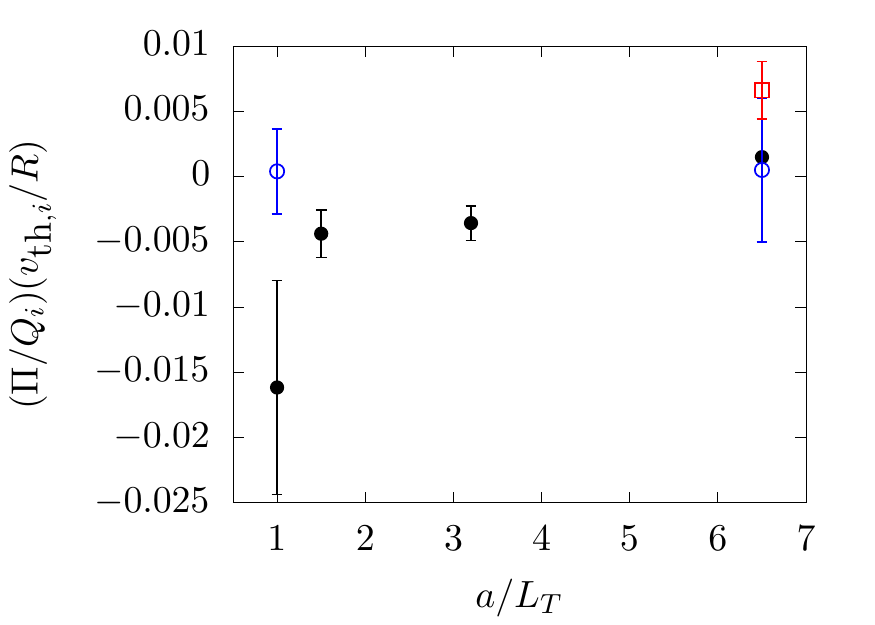}
     \includegraphics*[width=0.49\textwidth]{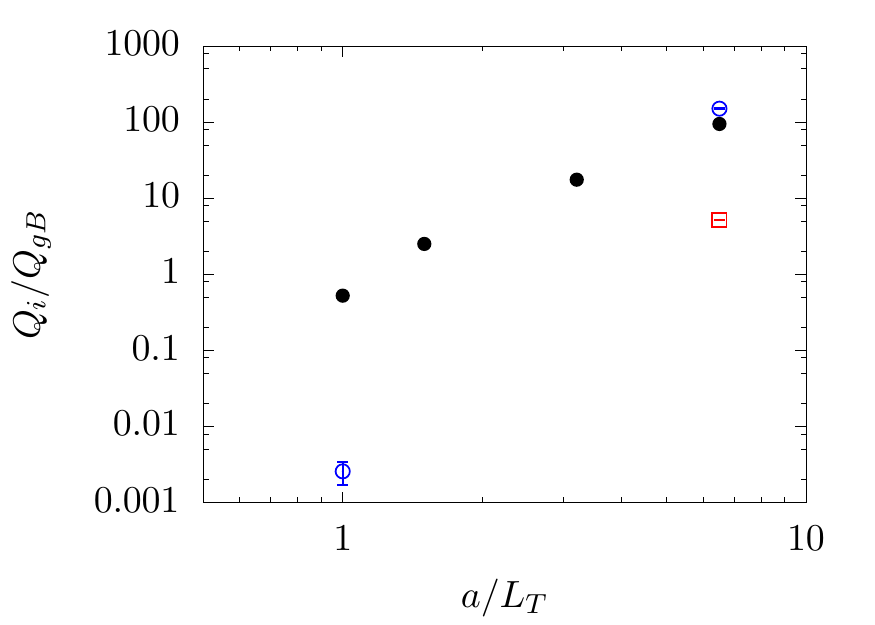}
  \end{center}
  \caption{(Left): Ratio of ion momentum flux $\Pi$ to ion heat flux $Q_{\rmi}$ as a function of $a/L_{T}$.  (Right): Normalized ion heat flux as a function of $a/L_{T}$, with $Q_{gB}\defeq n_{\rmi} T_{\rmi} \vthi (\rhoi/a)^2$. Blue open circles indicate cases where the radial component of the magnetic drift was artificially set to zero, and the red square is a case where both the radial and bi-normal components of the magnetic drift were artificially set to zero.  Error bars indicate statistical errors arising due to the finite interval used for the time average.}
  \label{fig:fluxes_vs_tprim}
\end{figure}

Time-averaged fluxes from nonlinear simulations run with $\rho_*=0.01$ and different $a/L_T$ values are given in Fig.~\ref{fig:fluxes_vs_tprim}.  After de-aliasing, the simulations included 128 Fourier modes in the radial wavenumber $k_x \defeq k_{\psi} r B_r / q$ and 22 Fourier modes in the bi-normal wavenumber $k_y\defeq k_{\alpha}B_r dr/d\psi$, with $\psi$ the poloidal flux.  The spacings in $k_y\rhoi$ and $k_x \rhoi$ were 0.05 and approximately 0.055 for all $a/L_T$ values except $a/L_T=6.5$, for which the spacings were approximately 0.033 and 0.037, respectively.  From the left panel of Fig.~\ref{fig:fluxes_vs_tprim}, we see that the ratio of ion momentum flux $\Pi$ to ion heat flux $Q_{\rmi}$ is approximately $\rho_*$ near marginal stability and decreases as the system gets further from marginal stability.  

The size of $\Pi/Q_{\rmi}$ near marginal stability is consistent with the scaling prediction given in~(\ref{eqn:pi2scaling}), but its decrease with increasing $a/L_T$ is not.  This is not entirely surprising given the discussion in Sec.~\ref{sec:reducedsymmetry} of an additional symmetry prohibiting momentum transport when the turbulence is concentrated at long wavelengths and when the radial magnetic drifts are unimportant.  Indeed, this is borne out by considering the behavior of the gyro-Bohm-normalized ion heat flux.  From the right panel of Fig.~\ref{fig:fluxes_vs_tprim}, we see that $Q_{\rmi}$ increases rapidly with distance from marginality, as expected.  Artificially removing the radial component of the magnetic drift results in more than an order of magnitude change in the heat flux near marginal stability, but only a few tens of percent change far above marginal stability.  When coupled with the fact that the turbulence peaks at wavelengths comparable to the poloidal Larmor radius far from marginality~\cite{barnesPRL11b}, this indicates that the additional symmetry discussed in Sec.~\ref{sec:reducedsymmetry} should be approximately satisfied.  From the left panel of Fig.~\ref{fig:fluxes_vs_tprim}, we see that the ratio $\Pi/Q_{\rmi}$ goes to zero (within error bars) when the radial magnetic drift is removed -- consistent with the presence of the additional symmetry of Sec.~\ref{sec:reducedsymmetry}.  This explains the small values of $\Pi/Q_{\rmi}$ for large $a/L_T$.

\begin{figure}[htpb]
  \begin{center}
    \includegraphics*[width=0.49\textwidth]{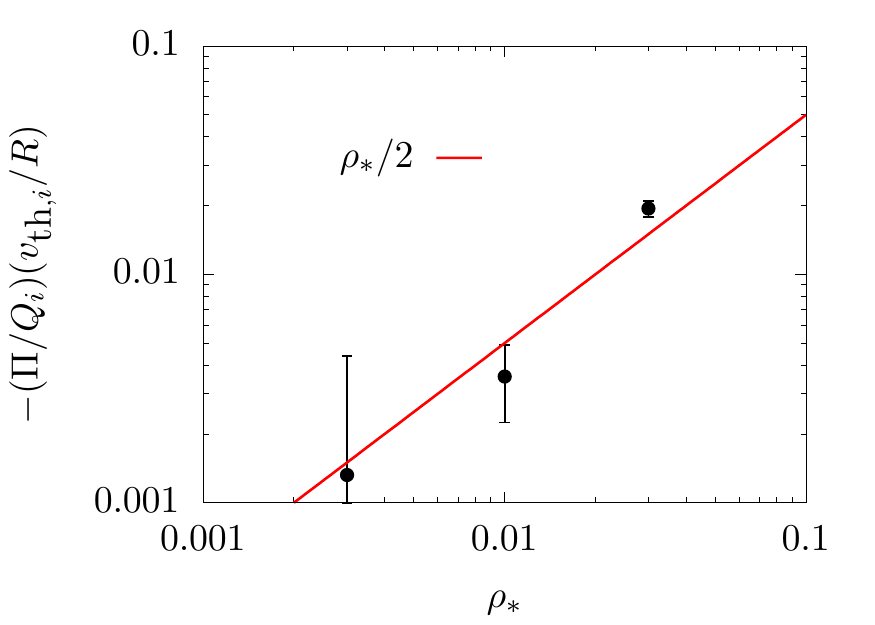}
  \end{center}
  \caption{Ratio of ion momentum flux $\Pi$ to ion heat flux $Q_{\rmi}$ as a function of $\rho_*=\rhoi/a$.  Error bars indicate statistical errors arising due to the finite interval used for the time average.}
  \label{fig:pioq_vs_rhostar}
\end{figure}

We consider the scaling of $\Pi/Q_{\rmi}$ with $\rho_*$ at fixed $a/L_T=3.2$ in Fig.~(\ref{fig:pioq_vs_rhostar}).  The data are consistent with a linear scaling in $\rho_*$, as expected for small $\rho_*$ given the perturbative framework in which we are working.

\section{Summary and discussion}\label{sec:conclusions}

The main results of the paper are encapsulated in Figs.~\ref{fig:fluxes_vs_tprim} and~\ref{fig:pioq_vs_rhostar}. They indicate, for the parameters chosen here, that the radial transport of toroidal angular momentum driven by turbulent parallel acceleration is similar to or smaller than $\rho_*$, regardless of the strength of the turbulence drive.  We argued in Sec.~\ref{sec:scalings} that this should be expected when turbulent eddies have a typical size of the ion gyroradius, as is the case near marginal stability.  Further from marginal stability, as turbulent eddies grow larger, the same scaling arguments predict that the ratio of momentum flux to heat flux should increase.  This discrepancy with simulation results is anticipated in Sec.~\ref{sec:reducedsymmetry} by noting that an additional, approximate symmetry of the gyrokinetic-Poisson system is satisfied when $\beta'$ is small, radial magnetic drifts are unimportant and turbulence is concentrated at long wavelengths.  These conditions are often satisfied far above marginal stability, as borne out by the data presented in Sec.~\ref{sec:results}.

To the extent that our results are applicable to a broader range of plasma parameters, our study implies that turbulent acceleration is unlikely to contribute significantly to intrinsic rotation.  This is because there are other symmetry-breaking mechanisms -- namely, neoclassical flows~\cite{barnesPRL13,leePoP14,leeNF14,leePPCF15,hornsbyNF17} and finite orbit width effects~\cite{parraNF11,parraPoP12,parraPPCF15} -- that have been found analytically and numerically to drive $\Pi/Q_{\rmi}$ that scales as $(B/B_p)\rho_*$.  As $B_p \ll B$ in most tokamaks, the associated momentum flux is likely an order of magnitude larger than the values obtained here.  There are, however, a couple of caveats to consider.  The scaling theory from Sec.~\ref{sec:scalings} was derived (and verified) under the assumption that turbulence is far from marginal.  As such, it is not clear from theoretical considerations alone if one should expect additional factors of $(B/B_p)$ appearing in the $\rho_*$ scaling of the momentum flux near marginal stability.  Of course, this study also only considered a single point in the parameter space; a broader range of parameters needs to be considered before a definitive statement about the importance of turbulent acceleration in generating intrinsic rotation can be made.

Finally, it is perhaps worth noting that the arguments used here to obtain the momentum flux scaling~(\ref{eqn:pi2scaling}) lead to an identical result for the momentum flux driven by the slow poloidal variation of turbulence and by radial profile variation~\cite{parraPPCF15}, both so-called 'global' effects.  The discussion from Sec.~\ref{sec:reducedsymmetry} also applies to these global effects, so that they too do not lead to momentum transport for a reduced system with no magnetic shear or radial magnetic drifts.  As such, the results reported here for turbulent acceleration may provide some insight as to the size and scaling of the momentum flux driven by global effects.

\ack

The authors acknowledge the use of ARCHER through the Plasma HEC Consortium EPSRC grant number EP/L000237/1 under project e281-gs2 and the use of the EUROfusion High Performance Computer (Marconi-Fusion) under project MULTEI.

\clearpage

\bibliographystyle{unsrt}

\end{document}